\documentclass[twocolumn,showpacs,aps,prl,letterpaper]{revtex4-1}
\usepackage{amsmath}
\usepackage{amssymb}
\usepackage{graphicx}
\usepackage{multirow}
\usepackage{float}
\usepackage{url}
\def\bra#1{\left\langle #1\right|}
\def\ket#1{\left| #1\right\rangle}
\def\beq{\begin{equation}}
\def\eeq{\end{equation}}
\def\bea{\begin{eqnarray}}
\def\eea{\end{eqnarray}}
\def\lam{\Lambda_c^+}
\def\lamS{\Lambda_c^+ \to B P}

\def\sss{\scriptscriptstyle}

\RequirePackage{xspace} \allowdisplaybreaks
\begin{document}

\title{\large \bfseries \boldmath Study of $CP$ violation in $\Lambda_c^+$ decay}
\author{Xian-Wei Kang$^{1,2}$}\email{kangxw@ihep.ac.cn}
\author{Hai-Bo Li$^2$}\email{lihb@ihep.ac.cn}
\author{Gong-Ru Lu$^1$}
\author{Alakabha Datta$^3$}\email{datta@phy.olemiss.edu}
\affiliation{$^1$Department of Physics, Henan Normal University,
Xinxiang 453007, China\\ $^2$Institute of High Energy Physics, PO
Box 918, Beijing  100049, China\\ $^3$ Department of Physics and
Astronomy,University of Mississippi, University, MS 38677, USA}

\begin{abstract}

In this paper, we study $CP$ violation in
 $\Lambda_c^+\to BP$ and $\Lambda_c^+\to BV$
decays, where $B,P$ and $V$ denote a light spin-$\frac{1}{2}$
baryon, pseudoscalar and a vector meson respectively. In these
processes the $T$ odd $CP$ violating triple-product (TP)
correlations are examined. The genuine $CP$ violating observables
which are composed of the helicity amplitudes occurring in the
angular distribution are constructed. Experimentally, by performing
a full angular analysis it is shown how one may extract the helicity
amplitudes and then obtain the TP asymmetries. We estimate the TP
asymmetries in $\lam$ decays to be negligible in the standard model
making these processes an excellent place to look for new physics.
Taking a two Higgs doublet model, as an example of new physics, we
show that large TP asymmetries are possible in these decays.
 Finally, we discuss how BES-III and Super
$\tau$-Charm experiments will be sensitive to these $CP$ violating
signals in $\Lambda_c^+$ decays.

\end{abstract}


\maketitle


Thanks to the effort that the $B$ factories have made over the last
decade, it has been confirmed that the Cabibbo-Kobayashi-Maskawa
(CKM) mechanism embedded in the standard model (SM) is the leading
source of $CP$ violation in the quark sector \cite{Cabibbo, KM}.
Especially for the $s$-quark and $b$-quark sectors, impressive
agreement between theory and experiment has been achieved
~\cite{Charles:2004jd,Bona:2006ah}. However, due to the small
contribution from the SM, $CP$ violation in the $c$-quark sector is
tiny making it an excellent place to look for new $CP$ violating
signals from new physics (NP). While $CP$ violation has not been
seen presently in the charm sector \cite{D mesons1,D mesons2,D
mesons3,D mesons4}, more precise measurements will be necessary to
test the SM predictions of $CP$ violation in this sector.
\cite{kang1,kang2,SM DDbar}. One can also search for new $CP$
violating phases by looking at baryon decays containing a charm
quark.  Studying $CP$ violation in $\Lambda_c^+$ baryon decays opens
up another front in the search for NP signals in the charm sector.
The $CP$ violation parameter $\mathcal{A}$ in $\Lambda_c^+$ decaying
to $\Lambda\pi^+$ has been measured by the FOCUS Collaboration in
2005 with the result of $\mathcal{A}= -0.07\pm0.19\pm0.12$
\cite{FOCUS}, where the errors are statistical and systematic,
respectively. The Particle Data Group (PDG) value is $\mathcal{A}$ =
$-0.07\,\pm$ 0.31 \cite{pdg}. Hence this measurement does not show
evidence of $CP$ violation and is therefore consistent with the SM.
However, the errors in the measurement are large and no definite
conclusion about the absence or presence of new $CP$ violating
phases can be made at this time.

Note that the parameter $\mathcal{A}$ is a $CP$ violating asymmetry
constructed out of the up-down asymmetry $\alpha$, and is a direct
$CP$ violating asymmetry. Therefore, $\mathcal{A} \sim \sin
{\delta}$, where $\delta$ is the strong phase difference between the
two amplitudes in the decay. Thus, if $\delta$ is small then
$\mathcal{A}$ may be small even in the presence of new $CP$
violating phases beyond the SM. There are theoretical estimates,
based on baryon chiral perturbation theory, that indicate that the
strong phases from $\Lambda-\pi$ rescattering are small at the $\Xi$
mass \cite{phaseshift1, phaseshift2}. It is not obvious that the
results of baryon chiral perturbation theory can be applied to
$\lam$ decays as the pion from the $\lam \to \Lambda \pi$ transition
has energy, $E_{\pi}=875$ MeV, which is close to the cut-off $ \sim
GeV$ for chiral perturbation theory. It is interesting to note that
if one applies the same equations in Ref.~\cite{phaseshift1,
phaseshift2} but evaluated at the $\lam$ mass then one gets small
strong phases of a few degrees. We start with
Ref.~\cite{phaseshift1}, where it was shown that the $S$-wave phase
shift vanishes in the leading order in baryon chiral perturbation
theory. The $P$-wave phase shift $\delta_1$
 can be expressed as,
\begin{eqnarray}\label{delta1}
\delta_1 &=& - \frac{(E_\pi^2-m_\pi^2)^{3/2}}{12\pi f_\pi^2}
[\frac{1}{4}\frac{g_{\Sigma\pi}^2}{E_\pi +
m_{\Sigma}-m_{\Lambda}}+\frac{3}{4}\frac{g_{\Sigma\Lambda}^2}{E_\pi+m_{\Lambda}-m_{\Sigma}}\nonumber\\
&&\qquad\qquad\qquad\qquad-\frac{4}{3}\frac{g_{\Sigma^*\Lambda}^2}{E_\pi+m_{\Sigma^*}-m_\Lambda}]
\end{eqnarray}
where $f_\pi \sim 132\,\, MeV$ is the pion decay constant and
$g_{\Sigma\Lambda}^2 \simeq 1.44$ is the strong
 $\Sigma$-$\Lambda \pi$ coupling. The strong coupling of the spin-$\frac{3}{2}$
isotriplet $\Sigma^*$ to $\Lambda \pi$ is determined to be
$g_{\Sigma^*\Lambda}^2 \simeq 1.49$ \cite{phaseshift1}. Evaluating
the phase shift at $E_\pi=875\,\,MeV$  we find
 $\delta_1
\simeq -0.05^\circ$. In Ref.~\cite{phaseshift2}, the $S$ wave phase
shift was generated in $\Lambda-\pi$ scattering, mainly through the
exchange of $\frac{1}{2}^- \Sigma(1750)$ (denoted by $\Sigma'$). In
this case the $S$-wave phase shift $\delta_0$ is given by,
\begin{equation}\label{delta0}
\delta_0 = - \frac{1}{2\pi}\frac{g_{\Sigma'\Lambda}^2}{f_\pi^2}
\frac{E_\pi^2(m_{\Sigma'}-m_\Lambda)\sqrt{E_\pi^2-m_\pi^2}}{E_\pi^2-(m_{\Sigma'}-m_{\Lambda})^2},
\end{equation}
In Eq.~\eqref{delta0}, the coupling parameter $g_{\Sigma'\Lambda}$
can be obtained from the branching ratio for $\Sigma(1750)\to
\Lambda\pi$. One has,
\begin{equation}
\Gamma[\Sigma(1750)\to \Lambda\pi] =
\frac{g_{\Sigma'\Lambda}^2}{4\pi f_\pi^2}(m_{\Sigma'}-m_{\Lambda})^2
\frac{p_\Lambda}{m_{\Sigma'}} (E_\Lambda+m_\Lambda),
\end{equation}
where $E_{\Lambda}$ and $p_{\Lambda}$ are the energy and momentum of
the $\Lambda$. The decay $\Sigma(1750)\to \Lambda\pi$ is not the
dominant decay mode for $\Sigma(1750)$ and the branching ratio for
this decay has not been measured though the decay has been
seen\cite{pdg}. For an estimate of $g_{\Sigma'\Lambda}$ we will
assume $BR[\Sigma(1750)\to \Lambda\pi] \simeq BR[\Sigma(1750)\to
\Sigma \pi] \le  $ 8 \% \cite{pdg} and setting the  width of
$\Sigma(1750)$ at $ 90\,\,MeV$ \cite{pdg} one obtains
$g_{\Sigma'\Lambda}\simeq 0.0058$. Substituting $g_{\Sigma^*\Lambda}
= 0.0058$ and $E_\pi=875\,\,MeV$ into Eq.~\eqref{delta0}, yields
$\delta_0 \le -3.4^\circ$.

Hence, chiral perturbation theory if applicable for $\lam$ decays,
predict small strong phases. The key point that we want to emphasize
here, is that $\mathcal{A}$ will not reveal new non-SM CP violating
phases if the strong phases are small and hence other $CP$ violating
signals that do not vanish with vanishing strong phases should also
be measured.

 In this Letter, we shall exploit the idea of ``triple-product (TP) correlation''
 to construct the $CP$ violating observables. These $CP$
violating quantities are proportional to $\cos{\delta}$ and so can
be large even with small strong phases. Hence these measurements are
complimentary to the direct $CP$ violation measurements mentioned
above. This type of $CP$ violation is not new and has been
considered previously
 in hyperon decays \cite{J.F.Donoghue1,J.F.Donoghue2}, in $B$ decays
 \cite{Valencia, Datta} and in other processes. In the $B$ system, Babar and Belle Collaborations have measured
TP asymmetries \cite{Babar1,Babar2,Belle} and these measurements
provide strong limits on the $CP$ violation originating from new
physics \cite{dattajhep}. The idea of TP correlation was also
considered in the $D$ meson sector \cite{YueLiangWu, FOCUS2, kang2}.
In Ref.~\cite{kang2} the authors proposed to measure TP asymmetries
in the $D$ meson decays at BES-III and they  estimated the errors in
the determination of these asymmetries at BES-III and in other
upcoming experiments.  At this point it is worth pointing out that
TP asymmetries in $D \to VV$ decays and $\Lambda_c$ decays could
probe different new physics because of the spin of the $\Lambda_c$.
As an example, consider NP of the type $L_{NP} \sim \bar{s} \gamma_A
c \bar{d} \gamma_B u$, where $\gamma_{A,B}= 1 , \gamma_5$. In the
factorization approximation this NP will not contribute to $D(D_s)
\to V V$ decays but will contribute to $\Lambda_c \to B P$ decays.
This is because, $<V|\bar{d} \gamma_B u|0>=<V|\bar{s} c|D(D_s)>=0$
while $<B|\bar{s} \gamma_A c |\Lambda_c>, <P|\bar{d} \gamma_5 u|0>
\ne 0 $. Of course such interactions can contribute to $ D(D_s) \to
VP$ or $D(D_s) \to PP$ decays but triple product asymmetries cannot
be constructed in such decays. It should be mentioned that
renormalization effects to the NP operator above can generate new
operator structures that will contribute to $D(D_s) \to VV$ decays
but these effects should be suppressed in general. In general, if NP
is detected in $D$ decays, T.P asymmetry measurement in $\Lambda_c$
decays can provide additional information about this NP.

An observable involving $\vec{v}_1\cdot(\vec{v}_2\times\vec{v}_3)$,
where each $v_i$ can be a spin or momentum, is called a TP
correlation. These TP's are odd under naive time reversal ($T$) and
hence constitute a $T$-odd $CP$ violating observable. One can define
an asymmetry quantity
\begin{equation}\label{A_T1}
A_T = \frac{N(\vec{v}_1\cdot\vec{v}_2\times\vec{v_3} >
0)-N(\vec{v}_1\cdot\vec{v}_2\times\vec{v}_3 < 0)}{N_{total}},
\end{equation}
where the subscript $T$ implies TP and $N$ denotes the number of
events. Equivalently one can define,
\begin{equation}\label{A_T2}
A_T  = \frac{\Gamma(\vec{v}_1\cdot\vec{v}_2\times\vec{v_3}
> 0)-\Gamma(\vec{v}_1\cdot\vec{v}_2\times\vec{v_3} < 0)}
{\Gamma(\vec{v}_1\cdot\vec{v}_2\times\vec{v_3}
> 0)+\Gamma(\vec{v}_1\cdot\vec{v}_2\times\vec{v_3} < 0)}\,.
\end{equation}
For its conjugate channel, the similar quantity $\overline{A}_T$ can
be defined in the same way. It should be noted that there is a
well-known technical complication: a non-zero TP correlation does
not necessarily imply $CP$ violation, since final state interactions
(FSI) can fake it, namely the strong phase can also produce non-zero
$A_T$ (or $\overline{A}_T$) even though the weak phases are zero
\cite{kang2,Valencia,Datta}. Yet comparing a TP correlation with its
measurement in a $CP$ conjugate transition allows one to distinguish
genuine $CP$ violation from FSI effects. One can define a true $CP$
violating asymmetry as,
\begin{equation}\label{A_T}
\mathcal{A_T} = \frac{1}{2}(A_T + \overline A_T),
\end{equation}
and hence a nonzero $\mathcal{A_T}$ is a  $CP$ violating signal.

We begin the first part of our analysis with $\Lambda_c^+\to BP$
decays, where $B$ denotes a light spin-$\frac{1}{2}$ baryon, $P$
denotes a pseudoscalar. The amplitude for the decay $\Lambda_c^+\to
BP$ can be written as \cite{S.Pakvasa,Datta2}
\begin{equation}\label{couplings a b}
\mathcal{M}_P \equiv A(\Lambda_c^+\to BP) = \bar{u}_{B}(a +
b\gamma_5)u_{\Lambda_c},
\end{equation}
where $a$ and $b$ are the parity violating and parity conserving
amplitudes for the decay. In the rest frame of the $\Lambda_c^+$ one
can reduce the above as,
\begin{equation}\label{amp_rest}
\mathcal{M}_P \equiv A(\Lambda_c^+\to BP) = \chi_{B}^{\dagger}(S + P
\vec{\sigma} \cdot \hat{q})\chi_{\Lambda_c},
\end{equation}
where $\hat{q}$ is a unit vector in the direction of the daughter
baryon, the $\chi$'s are the two component spinors and \bea S& = &
\sqrt{2m_{\Lambda_c}(E_B+m_B)}a \nonumber\\
P& = & -\sqrt{2m_{\Lambda_c}(E_B-m_B)}b. \ \label{sp} \eea
The absolute value squared of $\mathcal{M}_{P}$ can be obtained as,
\begin{eqnarray}\label{M_P square}
|\mathcal{M}_P|^2&=&(|a|^2-|b|^2)(m_B m_{\Lambda_c}+p_B\cdot
s_{\Lambda_c}p_{\Lambda_c}\cdot s_B\nonumber\\
&&\qquad\qquad\qquad -p_B\cdot p_{\Lambda_c}s_B\cdot
s_{\Lambda_c})\nonumber\\&&+(|a|^2+|b|^2)(p_B\cdot p_{\Lambda_c}-m_B
m_{\Lambda_c}s_B\cdot
s_{\Lambda_c})\nonumber\\
&&+2Re(ab^*)(m_{\Lambda_c}p_B\cdot
s_{\Lambda_c}-m_Bp_{\Lambda_c}\cdot s_B)\nonumber\\
&&+2Im(ab^*)\epsilon_{\mu\nu\rho\sigma}p_B^{\mu}s_B^{\nu}p_{\Lambda_c}^{\rho}s_{\Lambda_c}^{\sigma}.
\end{eqnarray}
Here the last term gives the TP which can be seen explicitly in the
rest frame of $\Lambda_c^+$, where it takes the form
$\vec{p}_B\cdot(\vec{s}_B\times \vec{s}_{\Lambda_c})$. It is
important to note that the TP involves  $s_{\lam}$, the spin of
$\lam$, which can be measured by observing its decay. One can, for
instance, look at the scattering process $e^+e^-\to
X(4630)\to\Lambda_c^+\overline{\Lambda}_c^-$ \cite{Lambda_c}. The
polarization of each $\Lambda_c$ can be measured in a way similar to
the one employed in the decay $J/\Psi\to\Lambda\overline\Lambda$
\cite{Majianping} by analyzing the decay of the final state
particles. The produced $\Lambda_c^+\overline{\Lambda}_c^-$ pair
will sequentially decay into a pair of conjugated channels
$\Lambda_c^+\to BP(V)$ and $\overline{\Lambda}_c^-\to \overline
B\overline P(\overline V)$. The angular distribution of the decay
products will contain information on the
$\Lambda_c^+\overline{\Lambda}_c^-$ polarizations and hence the TP
asymmetries.

An adequate formalism to calculate angular distributions is the
framework of helicity amplitudes, described for instance in
Refs.~\cite{Jackson,Jacob,Chung,helicity,kang3}. The decay chain is
described by the product of amplitudes corresponding to each
reaction. For a decay $X\to YZ$, we define polar angles
$(\theta_X,\phi_X)$ describing the momentum of particle $Y$ in the
rest frame of $X$ in a basis where the $z$-axis is defined by the
momentum of $X$ in the rest frame of its mother particle. The decay
amplitude depends on $(\theta_X,\phi_X)$ and is denoted by $A^{X\to
YZ}_{\lambda_Y\lambda_Z}$ where $\lambda_Y,\lambda_Z$ are the
helicities of the daughter hadrons. In the process $\Lambda_c^+\to
BP$, there are two helicity amplitudes $A_{\frac{1}{2},0}$ and
$A_{-\frac{1}{2},0}$. On the other hand, in Eqs.~\eqref{couplings a
b} and ~\eqref{M_P square}, $a$ and $b$ are the two relevant
coupling parameters for the decay. It can be easily shown, using
Eq~\eqref{amp_rest}, that the parameters $A_{\frac{1}{2},0},
A_{-\frac{1}{2},0}$ are linear combinations of $S$ and $P$ defined
in Eq.~\eqref{sp}. Consequently,
 one can get
\begin{equation}
Im(ab^*)\sim Im(A_{\frac{1}{2},0}A_{-\frac{1}{2},0}^{*}).
\end{equation}
Defining
\begin{equation}
A_T =
\frac{Im(A_{\frac{1}{2},0}A_{-\frac{1}{2},0}^*)}{\big|A_{\frac{1}{2},0}\big|^2
+ \big|A_{-\frac{1}{2},0}\big|^2}\, , \label{a_t}
\end{equation}
and
\begin{equation}
\overline A_T = \frac{Im(\overline A_{\frac{1}{2},0}\overline
A_{-\frac{1}{2},0}^*)}{\big|\overline A_{\frac{1}{2},0}\big|^2 +
\big|\overline A_{-\frac{1}{2},0}\big|^2}\, . \label{a_tbar}
\end{equation}
 the genuine $T$ violating signal, as discussed in Eq.~\eqref{A_T},
 reads \cite{Datta,kang2}
\begin{eqnarray}
\mathcal{A_T}=\frac{1}{2}\Big(\frac{Im(A_{\frac{1}{2},0}A_{-\frac{1}{2},0}^*)}{\big|A_{\frac{1}{2},0}\big|^2
+ \big|A_{-\frac{1}{2},0}\big|^2}+\frac{Im(\overline
A_{\frac{1}{2},0}\overline A_{-\frac{1}{2},0}^*)}{\big|\overline
A_{\frac{1}{2},0}\big|^2 + \big|\overline
A_{-\frac{1}{2},0}\big|^2}\Big)\, .
\end{eqnarray}

For the process $\Lambda_c^+\to \Lambda\pi^+\to(p\pi^-)\pi^+$, we
can construct the $CP$ violating observable
\begin{eqnarray}\label{genuine CPV}
\mathcal{A_T} =&&
\frac{1}{2}\bigg[\frac{Im(A_{\frac{1}{2},0}^{\Lambda_c\to \Lambda
\pi^+}A_{-\frac{1}{2},0}^{*\Lambda_c\to \Lambda \pi^+
})}{\Big|A_{\frac{1}{2},0}^{\Lambda_c\to \Lambda \pi^+}\Big|^2 +
\Big|A_{-\frac{1}{2},0}^{\Lambda_c\to \Lambda
\pi^+}\Big|^2}\nonumber\\&&\qquad+\frac{Im(\overline
A_{\frac{1}{2},0}^{\overline{\Lambda}_c\to \overline\Lambda
\pi^-}\overline A_{-\frac{1}{2},0}^{*\overline{\Lambda}_c\to
\overline\Lambda \pi^-})}{\Big|\overline
A_{\frac{1}{2},0}^{\overline{\Lambda}_c\to \overline\Lambda
\pi^-}\Big|^2 + \Big|\overline
A_{-\frac{1}{2},0}^{\overline{\Lambda}_c\to \overline\Lambda
\pi^-}\Big|^2}\bigg],
\end{eqnarray}
where the quantities involved can be extracted through the angular
distribution. Without loss of generality, for spin-up $\Lambda_c^+$
the angular distribution reads
\begin{eqnarray}\label{spin-up}
|M|^2_{(\Lambda\pi)}&\propto&\bigg[\cos^2\frac{\theta_{\Lambda_c}}{2}\cos^2\frac{\theta_{\Lambda}}{2}\Big|A^{\Lambda\to
p\pi^-}_{\frac{1}{2},0}\Big|^2\nonumber\\&&+\cos^2\frac{\theta_{\Lambda_c}}{2}\sin^2\frac{\theta_{\Lambda}}{2}\Big
|A^{\Lambda\to
p\pi^-}_{-\frac{1}{2},0}\Big|^2\bigg]\cdot\Big|A^{\Lambda_c\to
\Lambda\pi^+}_{\frac{1}{2},0}\Big|^2\nonumber\\
&&+\bigg[\sin^2\frac{\theta_{\Lambda_c}}{2}\sin^2\frac{\theta_{\Lambda}}{2}\Big|A^{\Lambda\to
p\pi^-}_{\frac{1}{2},0}\Big|^2\nonumber\\&&+\sin^2\frac{\theta_{\Lambda_c}}{2}\cos^2\frac{\theta_{\Lambda}}{2}\Big
|A^{\Lambda\to
p\pi^-}_{-\frac{1}{2},0}\Big|^2\bigg]\cdot\Big|A^{\Lambda_c\to
\Lambda\pi^+}_{-\frac{1}{2},0}\Big|^2\nonumber\\
&&-\frac{1}{2}\sin\theta_{\Lambda_c}\sin\theta_{\Lambda}\Big[\big|A^{\Lambda\to
p\pi^-}_{\frac{1}{2},0}\big|^2-\big|A^{\Lambda\to
p\pi^-}_{-\frac{1}{2},0}\big|^2\Big]\nonumber\\
&&\cdot\bigg[\cos\phi_{\Lambda}Re(A^{\Lambda_c\to
\Lambda\pi^+}_{\frac{1}{2},0}A^{*\Lambda_c\to
\Lambda\pi^+}_{-\frac{1}{2},0})-\nonumber\\
&&\qquad\sin\phi_{\Lambda}Im(A^{\Lambda_c\to
\Lambda\pi^+}_{\frac{1}{2},0}A^{*\Lambda_c\to
\Lambda\pi^+}_{-\frac{1}{2},0})\bigg],
\end{eqnarray}
and for a spin-down $\overline{\Lambda}_c^-$ it reads
\begin{eqnarray}\label{spin-down}
|\overline{M}|^2_{(\Lambda\pi)}&\propto&\bigg[\sin^2\frac{\theta_{\overline{\Lambda}_c}}{2}\cos^2\frac{\theta_{\overline\Lambda}}{2}\Big|A^{\overline\Lambda\to
\overline
p\pi^+}_{\frac{1}{2},0}\Big|^2\nonumber\\&&+\sin^2\frac{\theta_{\overline{\Lambda}_c}}{2}\sin^2\frac{\theta_{\overline\Lambda}}{2}\Big
|A^{\overline\Lambda\to \overline
p\pi^+}_{-\frac{1}{2},0}\Big|^2\bigg]\cdot\Big|A^{\overline{\Lambda}_c\to
\overline\Lambda\pi^-}_{\frac{1}{2},0}\Big|^2\nonumber\\
&&+\bigg[\cos^2\frac{\theta_{\overline{\Lambda}_c}}{2}\sin^2\frac{\theta_{\overline\Lambda}}{2}\Big|A^{\overline\Lambda\to
\overline
p\pi^+}_{\frac{1}{2},0}\Big|^2\nonumber\\&&+\cos^2\frac{\theta_{\overline{\Lambda}_c}}{2}\cos^2\frac{\theta_{\overline\Lambda}}{2}\Big
|A^{\overline\Lambda\to \overline
p\pi^+}_{-\frac{1}{2},0}\Big|^2\bigg]\cdot\Big|A^{\overline{\Lambda}_c\to
\overline\Lambda\pi^-}_{-\frac{1}{2},0}\Big|^2\nonumber\\
&&+\frac{1}{2}\sin\theta_{\overline{\Lambda}_c}\sin\theta_{\overline\Lambda}\Big[\big|A^{\overline\Lambda\to
\overline p\pi^+}_{\frac{1}{2},0}\big|^2-\big|A^{\overline\Lambda\to
\overline p\pi^+}_{-\frac{1}{2},0}\big|^2\Big]\nonumber\\
&&\cdot\bigg[\cos\phi_{\overline\Lambda}Re(A^{\overline{\Lambda}_c\to
\overline\Lambda\pi^-}_{\frac{1}{2},0}A^{*\overline{\Lambda}_c\to
\overline\Lambda\pi^-}_{-\frac{1}{2},0})-\nonumber\\
&&\qquad\sin\phi_{\overline\Lambda}Im(A^{\overline{\Lambda}_c\to
\overline\Lambda\pi^-}_{\frac{1}{2},0}A^{*\overline{\Lambda}_c\to
\overline\Lambda\pi^-}_{-\frac{1}{2},0})\bigg]\, .
\end{eqnarray}
 The first term in Eq.~\eqref{genuine
CPV} can be obtained from fitting to the angular dependence in
Eq.~\eqref{spin-up} and the second term can be obtained from fitting
to Eq.~\eqref{spin-down}. We also note that if the polarization of
the proton is known, each angular distribution in
Eq.~\eqref{spin-up} and Eq.~\eqref{spin-down} can be isolated into
two terms corresponding to the polarization states of the proton.

We next turn to the analysis of $\Lambda_c^+\to BV$ decays. The
general decay amplitude for this process can be written as
\cite{S.Pakvasa,Datta2}
\begin{eqnarray}
\mathcal{M}_V &=& A(\Lambda_c^+\to BV)\nonumber\\
              &=&\epsilon_{V\mu}^*\bar
u_{B}[(p_{\Lambda_c}^\mu+p_B^\mu)(a+b\gamma_5)\nonumber\\&&\qquad\qquad\qquad+\gamma^\mu(x+y\gamma_5)]u_{\Lambda_c}
\end{eqnarray}
where $\epsilon_{V\mu}$ is the polarization of the vector meson $V$,
$a, b, x$ and $y$ are coupling parameters. In the rest frame of
$\Lambda_c^+$, $p_V=(E_V, 0, 0, |\vec{p}_V|)$ and $p_B = (E_B, 0, 0,
-|\vec{p}_V|)$, thus the term
$\epsilon_{V\mu}^*(p_{\Lambda_c}^\mu+p_B^\mu)$ can be non-zero only
for the longitudinal polarized $V$. Evaluating $|\mathcal{M}_V|^2$
we get the relevant TP terms (refer to Ref.~\cite{Datta2}) as
\begin{eqnarray}\label{M_V square}
|\mathcal{M}_V|^2_{t.p.}& = &
2Im(ab^*)|\epsilon_V\cdot(p_{\Lambda_c}+p_B)|^2
                         \cdot\epsilon_{\mu\nu\rho\sigma}p_B^{\mu}s_B^{\nu}p_{\Lambda_c}^{\rho}s_{\Lambda_c}^{\sigma}\nonumber\\
                         &&+2Im(xy^*)\epsilon_{\mu\nu\rho\sigma}[\epsilon_V\cdot s_B p_B^{\mu} p_{\Lambda_c}^\nu s_{\Lambda_c}^{\rho}\epsilon_V^{\sigma}\nonumber\\
                         &&\qquad\qquad\qquad\quad-\epsilon_V\cdot p_B s_B^{\mu} p_{\Lambda_c}^\nu s_{\Lambda_c}^{\rho}\epsilon_V^{\sigma}\nonumber\\
                         &&\qquad\qquad\qquad\quad+\epsilon_V\cdot s_{\Lambda_c}p_B^{\mu}s_B^{\nu}\epsilon_V^{\rho}p_{\Lambda_c}^{\sigma}\nonumber\\
                         &&\qquad\qquad\qquad\quad-\epsilon_V\cdot p_{\Lambda_c}p_B^\mu s_B^\nu \epsilon_V^\rho s_{\Lambda_c}^\sigma]\nonumber\\
                         &&+2\epsilon_V\cdot(p_{\Lambda_c}+p_B)\epsilon_{\mu\nu\rho\sigma}\nonumber\\
                         &&\qquad\qquad\quad\cdot[Im(ax^*+by^*)p_B^{\mu}s_B^{\nu}p_{\Lambda_c}^{\rho}\epsilon_V^{\sigma}\nonumber\\
                         &&\qquad\qquad\quad+m_{\Lambda_c}Im(bx^*+ay^*)p_B^{\mu}s_B^{\nu}s_{\Lambda_c}^{\rho}\epsilon_V^{\sigma}\nonumber\\
                         &&\qquad\qquad\quad-Im(ax^*-by^*)p_B^{\mu}p_{\Lambda_c}^{\nu}s_{\Lambda_c}^{\rho}\epsilon_V^{\sigma}\nonumber\\
                         &&\qquad\qquad-m_B Im(ay^*-bx^*)s_B^{\mu}p_{\Lambda_c}^{\nu}s_{\Lambda_c}^{\rho}\epsilon_V^{\sigma}]
\end{eqnarray}
In Eq.~\eqref{M_V square}, aside from the first term, all the other
TP terms involve the polarization of the vector meson, and these TP
terms will vanish after summing over the polarizations of the vector
meson. In Eq.~\eqref{M_V square}, the first TP term survives only
for a longitudinal polarized $V$, and has the same structure as in
the $\lamS$ decay.
For the process
$\Lambda_c^+\to\Lambda\rho^+\to(p\pi^-)(\pi^+\pi^0)$, we can define
the TP asymmetry as
\begin{eqnarray}\label{genuine CPV2} \mathcal{A_T} =&&
\frac{1}{2}\bigg(\frac{Im(A_{\frac{1}{2},0}^{\Lambda_c\to\Lambda\rho^+}A_{-\frac{1}{2},0}^{*\Lambda_c\to\Lambda\rho^+})}{\Big|A_{\frac{1}{2},0}^{\Lambda_c\to\Lambda\rho^+}\Big|^2
+ \Big|A_{-\frac{1}{2},0}^{\Lambda_c\to\Lambda\rho^+}\Big|^2}\nonumber\\
&&\qquad+ \frac{Im(\overline
A_{\frac{1}{2},0}^{\overline{\Lambda}_c\to\overline\Lambda\rho^-}
A_{-\frac{1}{2},0}^{*\overline{\Lambda}_c\to \overline\Lambda
\rho^-})}{\Big|\overline A_{\frac{1}{2},0}^{\overline{\Lambda}_c\to
\overline\Lambda\rho^-}\Big|^2 + \Big|\overline
A_{-\frac{1}{2},0}^{\overline{\Lambda}_c\to
\overline\Lambda\rho^-}\Big|^2}\bigg)
\end{eqnarray}
The relevant terms(r.l.) occurring in the angular distribution are
listed below. For spin-up $\Lambda_c^+$ one has,
\begin{eqnarray}\label{spin up V}
&&|M|^2_{r.l.}\propto\Big[\cos^2\frac{\theta_{\Lambda_c}}{2}\cos^2\frac{\theta_{\Lambda}}{2}\cos^2\theta_\rho\big|A^{\Lambda\to
p\pi^-}_{\frac{1}{2},0}\big|^2\nonumber\\
&&\qquad+\cos^2\frac{\theta_{\Lambda_c}}{2}\sin^2\frac{\theta_{\Lambda}}{2}\cos^2\theta_\rho\big|A^{\Lambda\to
p\pi^-}_{-\frac{1}{2},0}\big|^2\Big]\cdot\Big|A^{\Lambda_c\to\Lambda\rho^+}_{\frac{1}{2},0}\Big|^2\nonumber\\
&&\qquad+\Big[\sin^2\frac{\theta_{\Lambda_c}}{2}\sin^2\frac{\theta_{\Lambda}}{2}\cos^2\theta_\rho\big|A^{\Lambda\to
p\pi^-}_{\frac{1}{2},0}\big|^2\nonumber\\
&&\qquad+\sin^2\frac{\theta_{\Lambda_c}}{2}\cos^2\frac{\theta_{\Lambda}}{2}\cos^2\theta_\rho\big|A^{\Lambda\to
p\pi^-}_{-\frac{1}{2},0}\big|^2\Big]\cdot\Big|A^{\Lambda_c\to\Lambda\rho^+}_{-\frac{1}{2},0}\Big|^2\nonumber\\
&&\qquad-\frac{1}{2}\sin\theta_{\Lambda_c}\sin\theta_{\Lambda}\cos^2\theta_\rho\Big[\big|A^{\Lambda\to
p\pi^-}_{\frac{1}{2},0}\big|^2-\big|A^{\Lambda\to
p\pi^-}_{-\frac{1}{2},0}\big|^2\Big]\nonumber\\
&&\qquad\cdot\Big[\cos\phi_{\Lambda}Re(A^{\Lambda_c\to
\Lambda\rho^+}_{\frac{1}{2},0}A^{*\Lambda_c\to
\Lambda\rho^+}_{-\frac{1}{2},0})-\nonumber\\
&&\qquad\qquad\qquad\sin\phi_{\Lambda}Im(A^{\Lambda_c\to
\Lambda\rho^+}_{\frac{1}{2},0}A^{*\Lambda_c\to
\Lambda\rho^+}_{-\frac{1}{2},0})\Big]\, ,
\end{eqnarray}
and for a spin-down $\overline\Lambda_c^-$
\begin{eqnarray}\label{spin down V}
&&|\overline{M}|^2_{r.l.}\propto\Big[\sin^2\frac{\theta_{\overline\Lambda_c}}{2}\cos^2\frac{\theta_{\overline\Lambda}}{2}\cos^2\theta_\rho\big|A^{\overline\Lambda\to
\overline p\pi^+}_{\frac{1}{2},0}\big|^2\nonumber\\
&&\qquad+\sin^2\frac{\theta_{\overline\Lambda_c}}{2}\sin^2\frac{\theta_{\overline\Lambda}}{2}\cos^2\theta_\rho\big|A^{\overline\Lambda\to
\overline p\pi^+}_{-\frac{1}{2},0}\big|^2\Big]\cdot\Big|A^{\overline\Lambda_c\to\overline\Lambda\rho^-}_{\frac{1}{2},0}\Big|^2\nonumber\\
&&\qquad+\Big[\cos^2\frac{\theta_{\overline\Lambda_c}}{2}\sin^2\frac{\theta_{\overline\Lambda}}{2}\cos^2\theta_\rho\big|A^{\overline\Lambda\to
\overline p\pi^+}_{\frac{1}{2},0}\big|^2\nonumber\\
&&\qquad+\cos^2\frac{\theta_{\overline\Lambda_c}}{2}\cos^2\frac{\theta_{\overline\Lambda}}{2}\cos^2\theta_\rho\big|A^{\overline\Lambda\to
\overline p\pi^+}_{-\frac{1}{2},0}\big|^2\Big]\cdot\Big|A^{\overline\Lambda_c\to\overline\Lambda\rho^-}_{-\frac{1}{2},0}\Big|^2\nonumber\\
&&\qquad+\frac{1}{2}\sin\theta_{\overline\Lambda_c}\sin\theta_{\overline\Lambda}\cos^2\theta_\rho\Big[\big|A^{\overline\Lambda\to
\overline p\pi^+}_{\frac{1}{2},0}\big|^2-\big|A^{\overline\Lambda\to
\overline p\pi^+}_{-\frac{1}{2},0}\big|^2\Big]\nonumber\\
&&\qquad\cdot\Big[\cos\phi_{\overline\Lambda}Re(A^{\overline\Lambda_c\to
\overline\Lambda\rho^-}_{\frac{1}{2},0}A^{*\overline\Lambda_c\to
\overline\Lambda\rho^-}_{-\frac{1}{2},0})-\nonumber\\
&&\qquad\qquad\qquad\quad\sin\phi_{\overline\Lambda}Im(A^{\overline\Lambda_c\to
\overline\Lambda\rho^-}_{\frac{1}{2},0}A^{*\overline\Lambda_c\to
\overline\Lambda\rho^-}_{-\frac{1}{2},0})\Big]\, .
\end{eqnarray}

  In experiment, the sizes and the relative phase between
$A_{\frac{1}{2},0}$'s and $A_{-\frac{1}{2},0}$'s can be extracted by
performing a full angular analysis given by Eqs.~\eqref{spin-up},
\eqref{spin-down}, \eqref{spin up V} and \eqref{spin down V}. These
expressions represent the main results of the paper. The measurement
of angular distributions and TP asymmetries can be done in a
reliable way with large data samples.

Note that, $\Lambda_c^+$ decays are different from $\Lambda_b$
decays, in the sense that in $\Lambda_b$ decay even the SM can give
large TP asymmetries \cite{Datta2,suubar,NP}. However  $CP$
violation in the charm sector is tiny in the SM, as indicated above,
and  hence any future non-zero signal of the $CP$ violation in
$\lam$ decays will be a signal of NP. We now estimate the size of
the TP asymmetry  in a model of NP. We begin with the SM where the
effective Hamiltonian for weak charm decays is given by
\cite{heff1,heff2} \beq H_{eff}^q = {G_F \over \protect \sqrt{2}}
[V_{cs}V^*_{ud}(c_1 O_{1}^q + c_2 O_{2}^q)] + h.c., \label{H_eff}
\eeq
where
\bea
O_1^q = {\bar s} \gamma_\mu (1-\gamma_5) c \, {\bar u} \gamma^\mu (1-\gamma_5) d ~, \nonumber\\
O_2^q = {\bar s}_\alpha \gamma_\mu (1-\gamma_5) c_\beta \, {\bar
u}_\alpha \gamma^\mu (1-\gamma_5) d_\beta, \ \label{H_effops} \eea
and $h.c.$ means Hermitian conjugate. We will use the Wilson's
coefficients at the charm scale as $c_1 = 1.27$, $c_2 = -0.53$
\cite{heff1,heff2}.

In the above we have neglected penguin contributions that are tiny
for the charm quark decay. In the absence of the penguin
contribution there is no TP asymmetry in the SM as there is only one
weak phase in the amplitude.

We now turn to a new physics model. We will consider a two Higgs
doublet model (2HDM) in which the decay $\lamS$ can get a
contribution through a charged Higgs exchange. The 2HDM is a simple
extension of the SM and is an effective low energy limit of many
extensions of the SM. We will provide a rough estimate of the T.P
asymmetry in this model. Our aim is to merely show that it is
possible for NP to generate a significant TP asymmetry in this
decay. The general Lagrangian for the $H^{\pm}ff^\prime$ interaction
is given by
\bea {\cal L}^{\sss 2HDM}& = & H^+ \left[\frac{y_c}{2}
\bar{c}(1-\gamma_5) s +
 \frac{y_u}{2} \bar{u}(1-\gamma_5) d + \frac{y_s}{2} \bar{c}(1+\gamma_5) s \right]
  \nonumber\\
 &&\quad +
 H^+\left[\frac{y_d}{2} \bar{u}(1+\gamma_5)\right]
d+ {\rm h.c.}\, \label{t2hdmInt} \eea
where $y_{c,s,u,d}$  are complex Yukawa couplings. There can also be
charged Higgs couplings for the $b$ quark which can cause deviations
from the SM in $B$ decays. However, in general the $b$ quark
coupling or the third generation couplings are  not related to the
couplings of the first two generations. Hence constraints on new
physics from $b$ quark decays do not apply to charm quark decays in
general.
 If the Higgs couples dominantly to the down quarks of the first two generation then we can assume $y_{c,u} << y_{s,d}$. Integrating out the heavy charged higgs leads to  the effective
 Hamiltonian for $\lamS$ transition
\bea H_{eff}^{NP} & = & {G_F \over \protect \sqrt{2}}
\frac{y_sy_d}{g^2} \frac{2 M_W^2}{m_{H_{+}}^2} \bar{s}(1-\gamma_5)c
\bar{u}(1+\gamma_5) d, \ \label{hnew} \eea where $g$ is the weak SM
coupling. We will now focus on the specific decay $\lam \to \Lambda
\pi^+$. In the presence of new physics we can write the amplitude
for $\lam \to \Lambda \pi^+$ as \bea
 A(\Lambda_c^+\to \Lambda \pi)& = & \bar{u}_{\Lambda}\left[a_{SM}(1+r_a) +
b_{SM}(1+r_b)\gamma_5)\right]u_{\Lambda_c}\nonumber\\
\label{npamp} \eea where $a_{SM}$ and $b_{SM}$ are the SM
contributions and $r_{a,b}$ are the ratios of the NP contributions
relative to the SM contributions. To estimate $r_{a,b}$ we will use
factorization and use the heavy quark limit for the charm quark.

To proceed with our calculations we use the following results for
the matrix elements, \bea \bra{\Lambda} \bar{s} (1-\gamma_5) c
\ket{\lam} &=&
\frac{q_{\mu}}{m_c}\bra{\Lambda} \bar{s} \gamma^\mu(1+\gamma_5) c \ket{\lam}\nonumber\\
\bra{\pi^+} \bar{u} \gamma^\mu(1-\gamma_5) d \ket{0} & = & i f_{\pi}
q^{\mu}
\nonumber\\
\bra{\pi^+} \bar{u} (1+\gamma_5) d \ket{0} & = & -i f_{\pi}
\frac{m_{\pi}^2}{m_u+m_d}, \ \label{me} \eea where
$q=p_{\Lambda_c}-p_{\Lambda}$, $f_{\pi}$ and $m_{\pi}$ are the pion
decay constant and its mass, $m_{u,d}$ are the up and down quark
masses. One can now compute $r_{a,b}$ in Eq.~\eqref{npamp} as, \bea
r_b & = & -r_a=r \nonumber\\
r & = & e^{i \phi}\frac{|y_sy_d|}{a_1 V_{cs}V_{ud}g^2} \frac{2
M_W^2}{m_{H_{+}}^2}\frac{m_{\pi}^2}{(m_u+m_d)m_c}, \ \label{r} \eea
where we have written $y_sy_d=|y_sy_d|e^{i \phi}$ with $\phi$ being
the new physics CP violating phase and $a_1=c_1 +c_2/N_c$. The
Yukawa couplings $y_{s,d}$ are unknown and can be $O(1)$. Assuming
the theory to be weakly coupled, we will take $y_{s,d} \sim g$ where
$g$ is the weak coupling. Using  $|y_sy_d| \sim g^2$, $a_1=0.94$,
$V_{cs}= 0.973$ \cite{pdg} and $V_{ud}=0.974$ \cite{pdg}  we find
$\frac{|y_sy_d|}{a_1 V_{cs}V_{ud}g^2} \sim 1.14 $ which gives $|r|
\approx 0.14$ for $m_{H^+}=300$ GeV. In our calculation we have
taken $m_c=1.4$ GeV, $m_d= 10$ MeV and $m_u= 5$ MeV \cite{pdg}. The
operators in the two Higgs doublet model can in principle be
constrained from  $D(D_s)$ decays. For instance, the new physics
operators can change the rate of $D(D_s)$ decays. However, as we
have shown above the size of NP is not that large and is at the
10-15\% level. Such size of NP are consistent with the measured $D$
decay rates because of hadronic uncertainties in the theoretical
calculations. For the same reason this size of NP is also consistent
with direct $CP$ measurements. As we have indicated in the paper TP
have advantages over other $CP$ measurements such as direct $CP$
asymmetry, if strong phases are small, and in general TP asymmetries
complement other $CP$ violation measurements. As indicated above the
scalar-pseudoscalar operators of the model will not contribute to
$D(D_s) \to VV$ at tree level but will contribute to TP asymmetries
in $\Lambda_c \to B P $ decays and hence can be constrained by
measurements in these decays.
Now we can estimate the TP asymmetry $ \sim 2|r| \sin{\phi} \frac{ 2
a_{SM}b_{SM}^*}{|a_{SM}|^2+|b_{SM}|^2} $. To a good approximation in
the SM, $a_{SM}=
\frac{m_{\Lambda_c}-m_{\Lambda}}{m_{\Lambda_c}+m_{\Lambda}}b_{SM}$.
One therefore obtains the TP asymmetry $  \sim 0.18 \sin {\phi}$ and
so a large TP is possible if $\sin{\phi} \sim 1$. Hence, it is
possible that future high luminosity experiments could provide
evidence for $CP$ violation in $\Lambda_c^+$ decays by exploiting
the method of $CP$ violating TP asymmetry. One can search for TP
asymmetries at BEPC-II/BES-III or future Super $\tau$-charm factory
with luminosity about 100 times as large as BEPC-II \cite{taucharm1,
taucharm2}.

 We now consider the potential sensitivities of the $CP$
violating observables $\mathcal{A}_T$ at BES-III and the Super
$\tau$-charm factory. From Eqs.~\eqref{A_T1} and \eqref{A_T2}, for a
small asymmetry, there is a general result that the error in
measurements is approximately estimated as $1/\sqrt{N_{obs}}$, where
$N_{obs}$ is the total number of events observed \cite
{kang1,kang2,Gronau,Li}. For the process $e^+e^-\to
X(4630)\to\Lambda_c^+\overline{\Lambda}_c^-$, taking the cross
section of $0.5 nb^{-1}$ into account \cite{Lambda_c},
 $2.5\times10^{6}\Lambda_c^+\overline{\Lambda}_c^-$ pairs
will be collected with an integrated luminosity of 5 $fb^{-1}$ at
$X(4630)$ peak for one year at BES-III. Table~\ref{tab1} lists some
promising $\Lambda_c^+\to BP$ channels to search for TP asymmetries.
 The expected statistical errors are estimated by
using $2.5\times 10^6 \Lambda_c^+\overline\Lambda_c^-$ pairs at
BES-III and $2.5\times 10^8 \Lambda_c^+\overline\Lambda_c^-$ pairs
at a Super $\tau$-charm factory. Table~\ref{tab2} shows the results
relevant to $\Lambda_c^+\to BV$ decays. The projected efficiencies
are estimated from the current status of BES-III and the branching
ratios are obtained from Ref.~\cite{pdg}.

\begin{widetext}
\begin{center}
\begin{table}[tbh]
\begin{center}
\begin{tabular}{ccccc}\hline\hline
$\multirow{2}*{BP}$ & \multirow{2}*{Br} & \multirow{2}*{Eff.($\epsilon$)} & Expected errors \\
                    &                   &                                 &   at BES-III ($\times10^{-2}$)   \\
\hline
$\Lambda\pi^+\to(p\pi^-)\pi^+$    &6.8$\times10^{-3}$    &0.82  &0.85  \\
$\Lambda K^+\to(p\pi^-)K^+$       &3.2$\times10^{-4}$    &0.75  &4.08  \\
$\Lambda(1520)\pi^+\to(p K^-)\pi^+$    &8.1$\times10^{-3}$  &0.75  &0.81  \\
$\Sigma^0\pi^+\to(\Lambda\gamma)\pi^+$ &$1.0\times10^{-2}$  &0.62  &0.80 \\
$\Sigma^0K^+\to(\Lambda\gamma)K^+$     &$4.0\times10^{-4}$  &0.56  &4.23 \\
$\Sigma^+\pi^0\to(p\pi^0)\pi^0$   &5.0$\times10^{-3}$       &0.60  &1.15 \\
$\Sigma^+\eta\to(p\pi^0)(\pi^+\pi^-\pi^0)$ &8.2$\times10^{-4}$ &0.52 &3.06\\
$\Xi^0K^+\to(\Lambda\pi^0)K^+$   &2.6$\times10^{-4}$           &0.57 &5.20\\
 \hline

\end{tabular}
\caption{The promising ($BP$) modes with  branching fractions,
efficiencies and expected errors on the TP asymmetries: the
corresponding expected errors are estimated by assuming $2.5\times
10^6 \Lambda_c^+\overline\Lambda_c^-$ pairs collected at BES-III
with one year luminosity.}\label{tab1}
\end{center}
\end{table}
\end{center}

\begin{center}
\begin{table}[tbh]
\begin{center}
\begin{tabular}{ccccc}\hline\hline
$\multirow{2}*{BV}$ & \multirow{2}*{Br} & \multirow{2}*{Eff.($\epsilon$)} & Expected errors \\
                    &                        &                                 &   at BES-III ($\times10^{-2}$)\\
\hline
$\Lambda \rho^+\to(p\pi^-)(\pi^+\pi^0)$              &3.2$\times10^{-2*}$   &0.65  &0.44  \\
$\Sigma(1385)^+\rho^0\to(\Lambda\pi^+)(\pi^+\pi^-)$  &2.4$\times10^{-3}$    &0.69  &1.55  \\
$\Sigma^+\rho^0\to(p\pi^0)(\pi^+\pi^-)$    &0.7$\times10^{-2*}$             &0.62  &0.96  \\
$\Sigma^+\omega\to(p\pi^0)(\pi^+\pi^-\pi^0)$    &1.4$\times10^{-2}$         &0.49  &0.76  \\
$\Sigma^+\phi\to(p\pi^0)(K^+K^-)$   &0.8$\times10^{-3}$                     &0.52  &3.10  \\
$\Sigma^+K^{*0}\to(p\pi^0)(K^-\pi^+)$ &0.7$\times10^{-3}$                   &0.57  &3.17  \\
\hline
\end{tabular}
\caption{The promising ($BV$) modes with branching fractions,
efficiencies and expected errors on the TP asymmetries: the
corresponding expected errors are estimated by assuming that
$2.5\times 10^6 \Lambda_c^+\overline\Lambda_c^-$ pairs collected at
BES-III with one year luminosity.}\label{tab2}
\end{center}
\end{table}
\end{center}
\end{widetext}

For the listed $BP$ and $BV$ modes in Table~\ref{tab1} and
Table~\ref{tab2}, the expected error in TP asymmetry measurement at
BES-III are estimated to be of the order of $\mathcal{O}(10^{-2})$.
At Super $\tau$-charm factory with $2.5\times10^8
\Lambda_c^+\overline\Lambda_c^-$ pairs, it will be reduced by one
order of magnitude. Hence the prospect of measuring TP asymmetries
in processes $\Lambda_c^+\to BP$ and $\Lambda_c^+\to BV$  at BES-III
and the Super $\tau$-charm factory are very promising. In Table
~\ref{tab2}, the branching ratios with asterisk have not been
measured, and we have set the branching fractions of the process
$\Lambda_c^+\to \Lambda\rho^+$ and $\Lambda_c^+\to \Sigma^+\rho^0$
to be at the upper limit values in PDG \cite{pdg}. Note that in
table~\ref{tab1} and \ref{tab2}, the estimated efficiencies are just
rough estimations according to the design of BEPC-II/BES-III. In the
future, careful measurements at BES-III of both the efficiencies and
branching fractions are suggested. A more realistic analysis would
require a likelihood fit to the full angular dependence of the
$\Lambda_c^+\to BP\to(B'P')P$ mode ($B',\,P'$ denote the daughter
baryon and pseudoscalar decay products of the parent particle $B$.)
 and of the $\Lambda_c^+\to BV\to(B''P'')(PP)$ mode ($B'',\,P''$
denote the daughter baryon and pseudoscalar decay products of the
parent particle $B$ and (PP) are the pseudoscalars from the decay of
the vector meson $V$.) Systematics will arise from
mis-reconstructions as some $B'P'$ can actually come from other
baryon resonances or non-resonant background contributions. In view
of the experimental realities, we expect that these systematics will
dominate the final result. Their precise estimates in the experiment
is beyond the scope of this paper.

 In conclusion, we studied the $CP$ violation in
$\Lambda_c^+\to BP$ and $\Lambda_c^+\to BV$ decay modes in which the
$T$-odd $CP$ violating TP correlations were examined. Here, $B,P$
and $V$ denotes a light spin-$\frac{1}{2}$ baryon, pseudoscalar and
a vector mesons,  respectively. We showed how the genuine $CP$
violating observable can be constructed and extracted from angular
distributions . These $CP$ violating observable depends on the
cosine of the strong phases, and if the strong phases are small,
they are potentially more sensitive to new $CP$ violating phases
beyond the SM than the direct $CP$ violating signals that depend on
the sine of the strong phases. We provided estimates of the TP
asymmetries in a model of NP and found that NP can produce large TP
asymmetries. Finally, we considered the potential sensitivities on
the $CP$ violating observable $\mathcal{A}_T$ at BES-III and at the
Super $\tau$-charm factory. Our numerical estimates showed that the
error in the measurements were very small and could reach the
magnitude of $\mathcal{O}(10^{-3})$. Hence, we concluded that the
prospect of measuring TP asymmetries in processes $\Lambda_c^+\to
BP$ and $\Lambda_c^+\to BV$  at BES-III and at the Super
$\tau$-charm factory were very promising.

{\bf Acknowledgement:}
This work is supported in part by the
National Natural Science Foundation of China under contracts Nos.
10521003, 10821063, 10835001, 10975047, 10979008, the National Key
Basic Research Program (973 by MOST) under Contract No.
2009CB825200, Knowledge Innovation Key Project of Chinese Academy of
Sciences under Contract No. KJCX2-YW-N29, the 100 Talents program of
CAS, and the Knowledge Innovation Project of CAS under contract Nos.
U-612 and U-530 (IHEP).

\end{document}